\documentclass[11pt,a4paper]{article}
\usepackage{cite,graphicx,dcolumn,bm,amsmath,amssymb}
\usepackage[english]{babel}

\begin{document}

\title{Conformations of entangled semiflexible polymers: \\ entropic
  trapping and transient non-equilibrium distributions}

\author{Hauke Hinsch and Erwin Frey \\ \small
Arnold Sommerfeld Center for Theoretical Physics and
  Center for NanoScience, \\ \small Department of Physics,
  Ludwig-Maximilians-Universit\"at M\"unchen, \\Ê\small Theresienstrasse 37,
  D-80333 M\"unchen, Germany}

\date{\today}

\maketitle

\begin{abstract}
The tube model is a central concept in polymer physics, and allows to reduce the complex many-filament problem of an entangled polymer solution to a single filament description. We investigate the probability distribution function of conformations of confinement tubes and single encaged filaments in entangled semiflexible polymer solution. Computer simulations are developed that mimic the actual dynamics of confined polymers in disordered systems with topological constraints on time scales above local equilibration but well below large scale rearrangement of the network. We observe the statistical distribution of curvatures and compare our results to recent experimental findings. Unexpectedly, the observed distributions show distinctive differences from free polymers even in the absence of excluded volume. Extensive simulations permit to attribute these features to entropic trapping in network void spaces.  The transient non-equilibrium distributions are shown to be a generic feature in  quenched-disorder systems on intermediate time scales.
\end{abstract} 
%

\thispagestyle{empty}

\section{Introduction}

Polymeric networks are not only versatile materials with a large variety of different mechanical properties but also provide interesting model systems for testing concepts of statistical mechanics. Such concepts usually aim to reduce the complicated many-body problem to a tractable single-polymer description as for instance in the well established tube model initiated by de Gennes \cite{gennes79} and Doi and Edwards \cite{doi86}. This reduction may be complicated by the fact that relevant dynamic processes for an individual polymer, its immediate surroundings and the complete network are each occurring on very different time scales. Furthermore, an additional length scale becomes important if the network is build of semiflexible polymers. 

A prominent model system that has recently been thoroughly studied for its relevance to biophysics \cite{bausch_kroy06} are filamentous actin (F-actin) solutions. F-actin strands at medium concentrations can form either chemically cross-linked networks in the presence of binding proteins or physically entangled networks for pure solutions with very different elastic properties \cite{hinner98,head03,wilhelm03,gardel04,heussinger06,wagner06}. These mechanical properties are experimental accessible by means of different rheological methods \cite{hinner98,amblard96} that mainly probe the collective properties resulting from the interaction of all network constituents on a macroscopic level. The fact that F-actin exceeds most synthetic polymers by length, furthermore permits to observe single network constituents on a microscopic level.  For instance, single polymers have been visualized to identify tube-like regions along which filaments reptate \cite{perkins94,kas94}. The availability of experimental observation on very different length scales allows to address one of the central questions of polymer physics, how the individual constituents and their behavior collectively determines the macroscopic properties of the polymeric material.

The challenge one is facing in developing a theory for the macroscopic properties of any material that builds on the underlying microscopic physics lies in finding a suitable simplification to describe the large number of network constituents without losing emergent properties.  In entangled networks, the only interactions present are of topological nature, as polymers can effortlessly slide past each other but are not allowed to cross.  These topological constraints mutually restrict the accessible configuration space of the polymers on intermediate time scales. To account for these entanglement constraints Edwards and de Gennes have introduced the tube model. This surprisingly simple approach successfully reduces the many-polymer problem of a network to a coarse-grained description. The suppression of transverse undulations of a test polymer by the surrounding polymers is mimicked by a hypothetical tube that confines the encaged polymer to a narrow cylindrical pore.  This tube follows the average path of the test polymer and its profile is usually described by a harmonic potential. The average strength of this potential and thereby the dimension of the tube is determined by the local density of the network. Semenov derived a scaling law for the dependence of the tube width on monomer concentration by using the assumption that fluctuating filaments explore non-overlapping regions of space \cite{semenov86} and Odijk introduced the deflection length as the length scale between collisions of the probe filament with the tube walls \cite{odijk83}. These scaling laws describing microscopic properties of single polymers in a network are the basis of further theoretical predictions for emerging macroscopic properties. For example, the confinement free energy of the filament inside the tube allows to connect the tube width to mechanical properties of the network 
\cite{hinner98,mackintosh95,isambert96}. More recently, even nonlinear extension of the standard tube model have been proposed \cite{fernandez09}.

In this work we go beyond the description of the tube in terms of its average size. We analyze the conformations of tube contours by focussing on the curvature distributions of tube contours and confined polymers. These quantities can provide useful information about equilibration processes and dynamics of confined polymers and are at the foundation of reptation theories \cite{doi85}. While it is usually assumed that the ensemble of confinement tubes or confined polymers qualitatively shows the same conformation statistics as free polymers, we challenge this assumption.

We will proceed as follows: in Section \ref{sec:system} the system
under investigation is defined and all relevant length and time scales
are discussed. We identify the characteristic energy distributions in
the tube model and present the description in two spatial dimensions.
In Section \ref{sec:simulations} we present our approach to simulate
the complete network by a probe filament in a two-dimensional array of
obstacles. We pay special attention to the detailed nature of the
Monte-Carlo moves used before we present results for the curvature
distribution of tubes and filaments and compare them to experiments.
As these results seem to disagree with standard concepts of
statistical mechanics at a first glance, we devote Section
\ref{sec:theory} to a thorough analysis of the underlying physics and
explain the cause of the surprising results. We corroborate this
explanation by further simulations before we concluding in 
Section \ref{sec:conclusion}.

\section{Tube Model} 
\label{sec:system}

We consider entangled solutions of semiflexible polymers where
chemical bonds by cross-linking proteins are ruled out. While F-actin
is a prominent example of this class of biopolymers and has a strong
record of experimental data available, our work is also
applicable to other semi-flexible polymer solutions in a comparable
regime where a description in the realm of the tube model is
justified \cite{wagner07}. As binding by cross-linking proteins is
ruled out, the only inter-polymer interactions are topological
constraints since network constituents
cannot mutually cross each other. The polymers are considered to be
mathematical lines as their thickness is negligible \cite{holmes90} and no
noteworthy long-ranged attractive or repulsive interactions exists in typical experimental situations.
Consequently the system has no excluded volume. In general, F-action
solutions are polydisperse with a mean filament length $L \approx 22
\mu$m \cite{kaufmann92}. The detailed length distribution, however, is
highly variable for different preparations \cite{kas96,kawamura70} and
we will consider monodispersity in the following. With a persistence
length $l_{\rm p} \approx 17 \mu$m \cite{gittes93,goff02} comparable
to its length F-actin is a typical semi-flexible polymer. The
polymer's bending stiffness $\kappa$ is related to the persistence length
as $l_{\rm p}=\kappa/k_{\rm B} T$ and each polymer's configuration ${\bf
  r}(s)$ is parameterized by the arc length $s$. A free polymer then
is described in the worm-like chain model \cite{kratky49,saito67} by
the Hamiltonian
\begin{equation} \label{eq:hamilton_free}
H(\kappa)=\frac{\kappa}{2} \int_0^L ds  \left( \frac{\partial^2 {\bf
    r}(s)}{\partial s^2} \right)^2 \;,
\end{equation} 
where the second derivative of ${\bf r}(s)$ is the local curvature
$\mathcal C$ at arc-length $s$. As a consequence, the distribution of local
curvatures of a free polymer is
\begin{equation} \label{eq:distr_free}
P(\mathcal C) \propto \exp \left( -\frac{1}{k_{\rm B} T}
  H(\kappa,\mathcal C) \right) 
\propto \exp \left( -l_{\rm p} \mathcal C^2 \right) 
\end{equation}
and the resulting Gaussian distribution's width decreases with
increasing persistence length of the polymers. 

The density $\nu$ of a
network of these polymers is given by the number of polymers of length
$L$ per unit volume. At a concentration of $c=0.5$~mg/ml corresponding
to $\nu \approx 1 \mu m^{-3}$ \cite{schmidt89} the average distance to
the next neighbor is given by the mesh mesh size $\xi \approx (\nu
L)^{-1/2} \approx 0.2 \mu$m and therefore much smaller then polymer
length and persistence length, $\xi \ll l_{\rm p}$. Due to this ratio
of length scales it is guaranteed, that a specific polymer will not
deviate far from its average contour and it is highly unlikely to fold
back onto itself. Thus it is feasible to model the combined effect of
all neighboring filaments of an arbitrary probe polymer by a
hypothetical tube potential. The tube potential has a harmonic
profile as observed in experiments and simulations
\cite{dichtl99,hinsch07}. Due to the disorder in the network the tube
diameter and thus the local strength of the tube potential vary along
the contour \cite{romanowska09}. The tube is conventionally described
by a potential strength $\gamma(s)$ that is parameterized by the arc
length $s$ along the tube backbone or tube contour given by the
potential's minimum in space. If we denote this tube backbone by ${\bf
  r}_0(s)$, the resulting energy becomes the sum of the bending energy
of the polymer and its confinement by a harmonic potential around the
tube backbone
\begin{equation} \label{eq:hamilton} H(\gamma,\kappa)=\int_0^L ds
  \left[\frac{\kappa}{2} \left( \frac{\partial^2 {\bf r}(s)}{\partial
        s^2} \right)^2+\frac{\gamma}{2} \left[ {\bf r}(s)-{\bf r}_0(s)
    \right]^2 \right] \;.
\end{equation} 
In contrast to a free polymer this equation does not allow to infer a
simple distribution of curvatures as in Eq.  \ref{eq:distr_free}
because the second term causes an additional dependence on the tube
contour.  This term causes a confinement around the minimum of the
potential that is given by the tube backbone as explained above. If
for instance the backbone is already strongly bend and the confinement
potential is sufficiently strong, high curvatures are more likely than
for a free polymer. Obviously the distribution of curvatures of any
probe polymer sensitively depends on the actual form of the tube to
which it is confined. Furthermore, the conformations of the tube backbone is itself a
statistically distributed quantity.  Finally, it is
important to state that a tube is well-defined only up to some
intermediate time scale. Therefore, in networks of non cross-linked
polymers the tube model cannot be used as an equilibrium concept
without further thought. The confinement tube is defined as the space
accessible to an encaged polymer in an environment of neighboring
filaments before large scale reconstruction of the network changes
this environment.  The tube picture is thus a valid description as
soon as the polymer experiences topological interaction with its
neighbors and as long as these are not remodelled by large scale
dynamics. The first time scale can be estimated from dynamic light
scattering as the point were a cross-over from free filament to
restricted dynamics sets in at about 10 ms \cite{semmrich07}. An estimate 
for the time scale of remodelling can be obtained from the
time it takes the probe filament to leave its initial tube. For
unstabilized actin filaments this process is dominated by treadmilling
occurring at an approximate rate of 2 $\mu m$ per hour \cite{selve86}.
Stabilized actin filaments where treadmilling is abolished by
phalloidin can only reptate out of their tubes at much slower rates.
Reptation rates in this case have been estimated from experiments
\cite{dichtl02,keller03} to be as long as several days for a 10 $\mu
m$ long filament.

\begin{figure}[htb]
\begin{center}
\includegraphics[width=0.6\columnwidth]{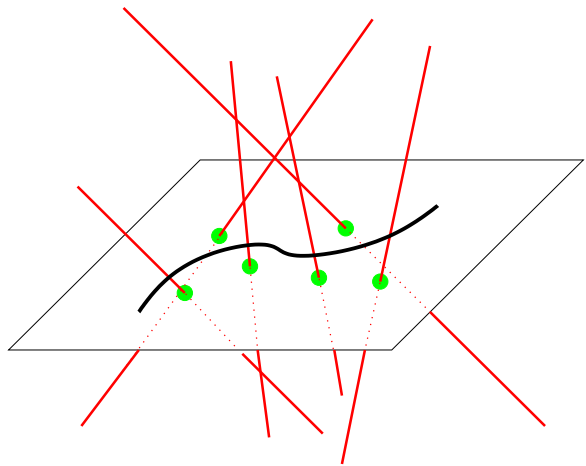}
\caption{In a two-dimensional cut through the network, as e.g. in the
  focal plane of a microscope, neighboring polymers reduce to
  point-like obstacles. \label{fig:intersection}}
  \end{center}
\end{figure}

As explained above, a well established scaling law for the average
tube diameter has been derived by Semenov. Recently, also theories
providing absolute values have been proposed \cite{hinsch07,morse01}
and confirmed experimentally \cite{romanowska09}. In analyzing
experimental data it is important to keep in mind that fluorescence microscopy only provides an observation of an
effectively two-dimensional focal plane. For the tube diameter 
this implies that only fluctuations in one Cartesian
component are measured. In the focal plane the system can be
considered as a probe filament surrounded by fluctuating point obstacles
that represent the cuts of neighboring polymers through the the plane
of observation as depicted in Fig. \ref{fig:intersection}.

\section{Monte-Carlo Simulations} \label{sec:simulations}

Motivated by this point of view, we have developed a Monte-Carlo
simulation with the standard Hastings-Metro\-polis algorithm
\cite{hastings70} to observe curvature distribution functions of a
single probe filament in a two-dimensional plane of point-like
obstacles. This does not only provide ready comparability with
experiments, but is also a valid simplification of the
three-dimensional problem since transverse undulations in different
components can be assumed to be independent. As the surrounding
network constituents are very thin and nearly straight due to their
large persistence length it is justified to represent them by
point-like obstacles. Those obstacles undergo themselves transversal
fluctuations that have to exhibit on average the same characteristics
as the fluctuations of the probe filament for reasons of
self-consistency. As the latter were found to be harmonic, the same
must hold for the in-plane fluctuations of an obstacle polymer that
cuts perpendicularly through the simulation plane. If the cutting
angle is tilted the harmonic profile is distorted corresponding to
different fluctuation strength for the two in-plane components. While
the average tube diameter must equal the fluctuation strength averaged
over all obstacles, the distribution of these two quantities may be
broad \cite{hinsch07}. It was found, however, that the simulation
results are rather insensitive to these parameters. The simulation is
thus an adequate and self-consistent description of the physical
problem of a entangled network of semi-flexible polymers, if the
parameters obstacle density $\rho$ and obstacle fluctuation potential
strength $\gamma$ are chosen to represent the corresponding polymer
density $\nu$ and resulting tube diameter $L_\perp$. These parameters
have been shown \cite{hinsch07} to be:
\begin{equation}
\rho=\frac{2}{\pi}\nu L \;, \qquad \gamma \approx 4.18 (\nu L)^{3/5}
l_p^{1/5} \;.
\end{equation}

The probe filament of length $L$ is initially placed in a straight
configuration onto the plane of observation and is represented by a
sequence of $N$ connected segments $i$ with orientation ${\bf t}_i$. Due
to inextensibility the segments are of fixed length $L/N$. In a first
step the filament is allowed to relax on the plane without the
presence of any obstacles.  The relaxation is performed with respect
to an Hamiltonian :
\begin{equation} \label{eq:htan}
H(\{{\bf t}_i\})=- J \sum_{i=1}^{N-1} {\bf t}_i  {\bf t}_{i+1} \;,
\end{equation}
where in two dimensions the relation between persistence length and
$J$ is given as $l_p/L=-[N \ln(I_1(J)/I_0(J))]^{-1}$ with $I_n$ the
modified Bessel functions of first kind \cite{fisher64}. After
equilibration with respect to the Hamiltonian (\ref{eq:htan}) the probe
filament features the bending distribution of a free polymer. Now
the obstacle fluctuation centers $p^0_j$ are fixed to random positions
of the simulation plane. While these centers remain fixed for the
course of the simulation, the positions of the point obstacles $p_j$ themselves -
initially placed at $p_j=p^0_j$ - are allowed to move in a harmonic
potential $U=\frac{1}{2} \gamma_j (p_j-p^0_j)^2$.  Their motion is not
only governed by this potential but also by the constraint that they must
not cross the probe filament and they remain on that side of the probe
filament where they had been initially placed \footnote{We also
  performed simulations where the topology of the obstacles, i.e. the
  side of the filament they are constrained to, is only determined
  after they are allowed to relax away from their fluctuation center.
  Thereby initial conditions with the probe filament lying between an
  obstacle and its fluctuation center become possible. The presence or
  absence of these ``misfit''-configurations does not change our
  results.}.  Naturally, the same constraint also holds for the probe
filament where every move, that would lead to a configuration
where an obstacle point had switched sides, is rejected.  While the
motion of the point obstacles is straightforward, we will discuss the
moves of the probe filament in more detail in the following section.

\subsection{Dynamic Trial Moves}

In the construction of trial moves our intention has been to find a
set of moves that mimics the underlying local dynamics in the physical
system as closely as possible. To this end it is of particular
importance to keep the relevant time scales in mind. Therefore, our
choice of moves describes the effect of the underlying physical forces
and dynamics on a probe filament for times well below large scale
rearrangement of the network. These have to include transverse
undulations, exploration of void spaces along the contour and
small-scale reptation and breathing, while the effects of large scale
reptation like annihilation and creation of obstacle points remain
impossible. In the following, we will thoroughly explain the moves
used in our Monte-Carlo simulations.  As shown below, the resulting
data for conformations and distribution functions sensitively depends
on their nature.

\begin{figure}[htbp]
\begin{center}
\includegraphics[width=0.6\columnwidth]{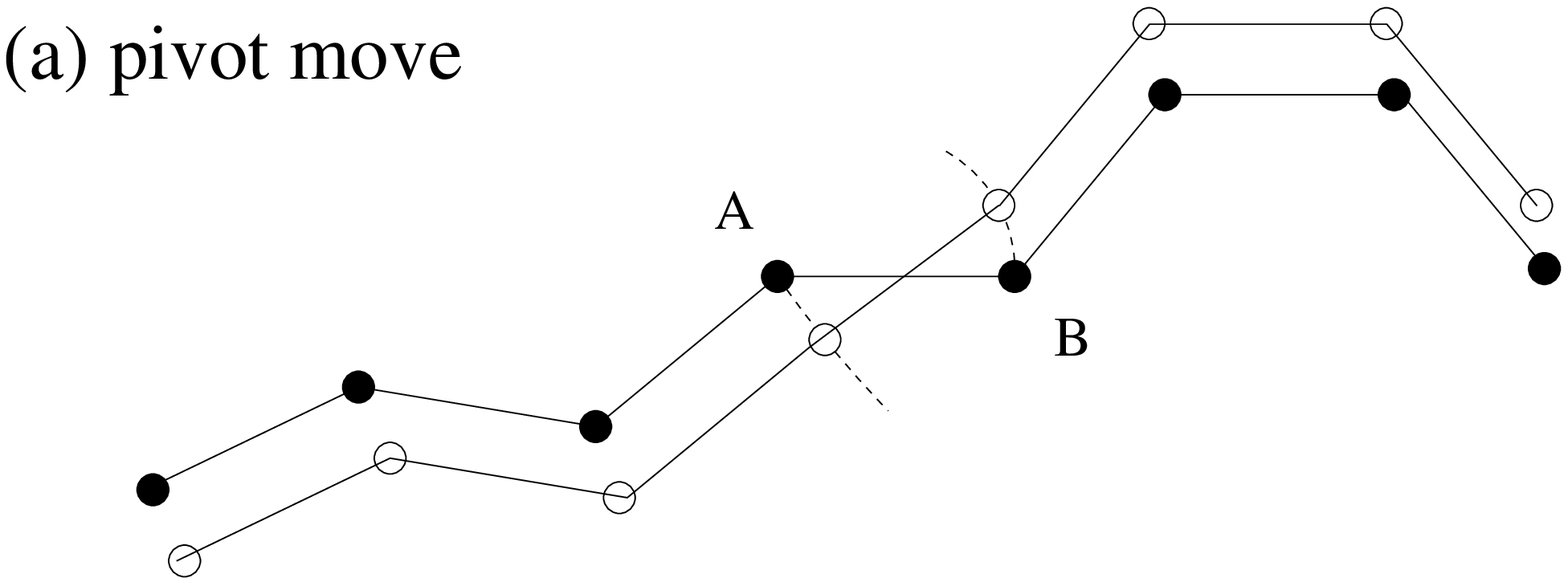}
\hspace{.5cm}
\includegraphics[width=0.6\columnwidth]{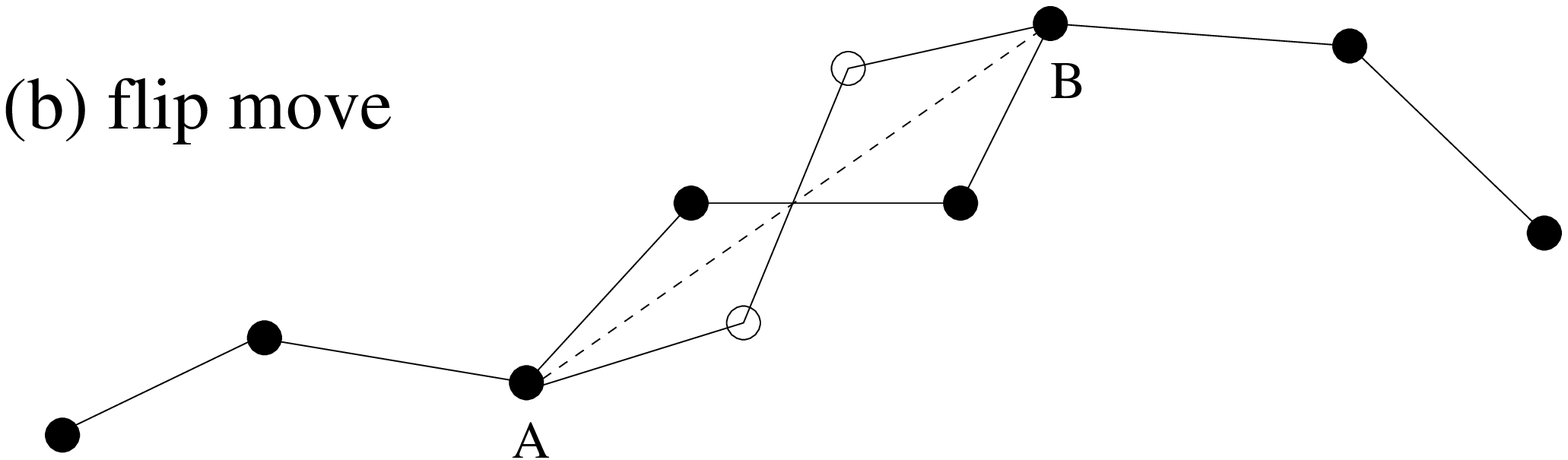}
\hspace{.1cm}
\includegraphics[width=0.6\columnwidth]{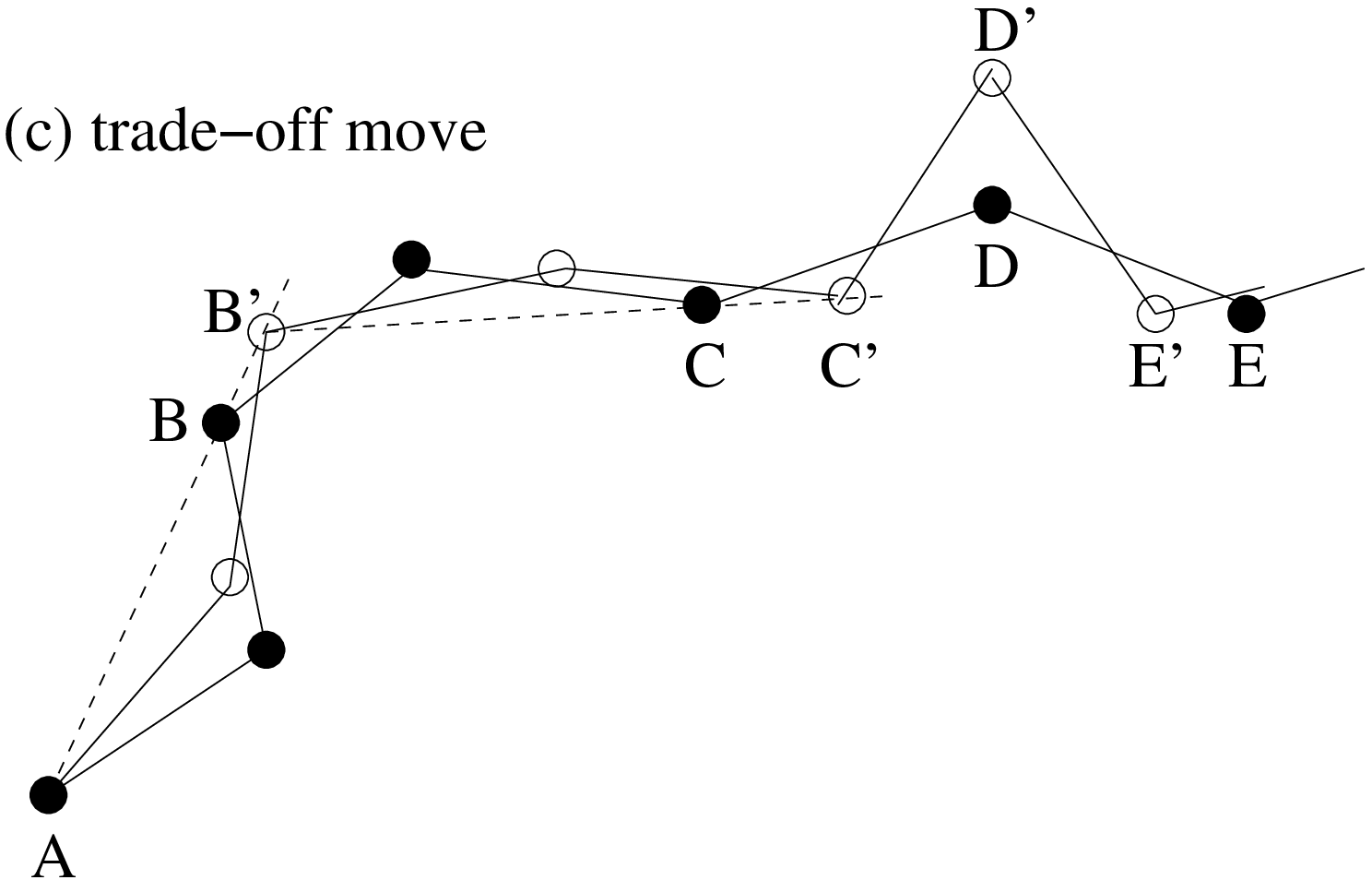}
\hspace{.1cm}
\includegraphics[width=0.6\columnwidth]{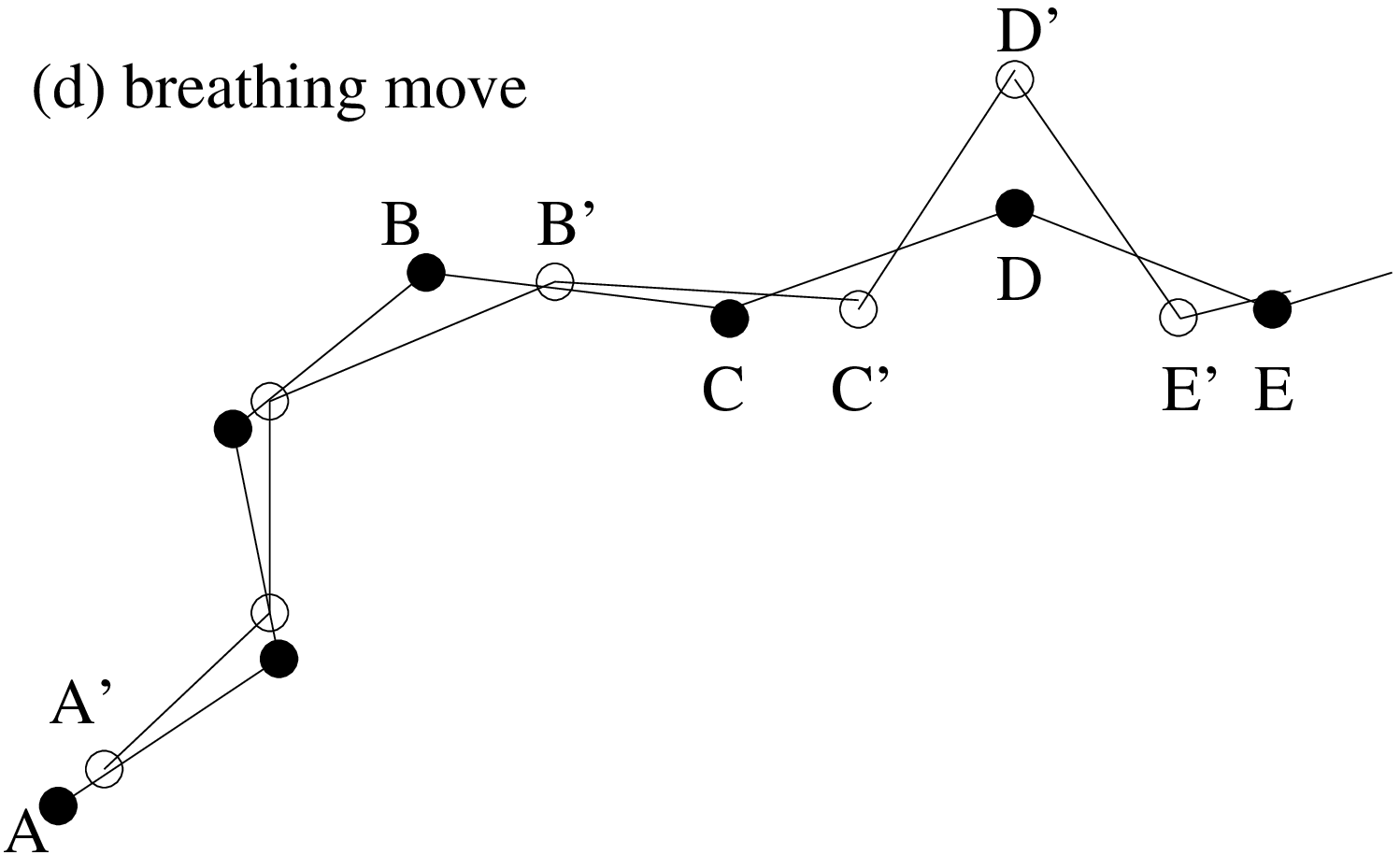}
\end{center}
\caption{The different moves performed during the simulations. Full
  circles denote the original positions and open circles the new
  configuration: {\it (a)} The pivot move changes two angles and all
  positions. {\it (b)} The flip move changes all angles and positions
  between A and B. {\it (c)} The ``trade-off'' move changes the
  curvature of the section CE in a trade-off with the adjacent section
  to both sides (only l.h.s section AC depicted here). {\it (d)} The
  ``breathing'' move changes all positions and angles and axially
   moves the polymer's ends. \label{fig:moves}}
\end{figure}

We use four different moves, that change the polymer configuration on
different levels ranging from local change in only one tangent up to a
global manipulation of the complete polymer contour. The classical
random pivot move (see Fig. \ref{fig:moves} (a)) chooses a random
tangent AB and rotates it by a small random angle.  This changes the
two bending angles at A and B but leaves all other angles invariant.
However, all positions along the polymer contour are modified -
predominantly in a direction transverse to the contour. Another move
employed is a flip move (see Fig. \ref{fig:moves} (b)).  Here two
beads A and B separated by a random number of segments are picked and
all points are mirrored along the axis connecting A and B. This can be
seen as the two-dimensional analogon to a crankshaft move. The flip
move leaves the ends of the polymer unaffected. These two rather
common moves are known to effectively explore the available phase
space of free polymers or even of polymers in pore-like potentials.
However, they lack the ability to mimic the motion of a polymer into
the void spaces between obstacles along the walls of the hypothetical
tube.

To this end we use novel moves that are depicted in Fig.
\ref{fig:moves} (c) and (d). They simulate the exploration of a local
void space, i.e. a part where the ``tube'' formed by the fluctuating
obstacles is rather large. Due to the negligible longitudinal
extensibility of semi-flexible polymers the motion of the polymer into
this void space is obviously possible only if the other parts of the
polymer are retracted or straightened out. Let us first introduce the
move that performs the latter and is illustrated in \ref{fig:moves}
(c). The additional length that is needed to enable the protrusion
into a void space is obtained from undulations in adjacent parts of
the polymer. A stronger bending in one part is made possible by weaker
bending in other parts and we therefore choose the label ``trade-off''
move. Specifically, the move is conducted by randomly choosing two
points C and E, that are separated by an even distance of segments (in
our simulations we apply two different moves where this distance is
either two or four).  This is the region of the polymer that should
explore the void space.  Furthermore, to the left and right of this
region two more points are chosen that enclose the regions that will
be straightened out to compensate for the contour length drawn into
the void. At the left hand site this is for example point A that is
separated from C by an again even number of tangents (in our
simulations two to ten). We restrict our illustration to the left hand
site, but the same process also applies to a comparable region on the
right hand side that is not shown. In addition to a random number
$R_{\mathrm{loc}}$ to choose the location of the section $CE$ along
the contour, we draw another random number $R_{\delta}$ that
quantifies the extend of change in bending. Our goal is now to
construct a algorithm that straightens out the region AC in order to
enable the region CE to realize a stronger bending. In a first step
the vector AB is prolonged to AB' by adding a random small amount
$\delta$. The new point between A and B is chosen in a way to keep
tangent length conserved.  In a next step B'C is prolonged to B'C' by
adding the same $\delta$ and so on. The same process is performed on
the right hand side tail. This results in two points C' and E' that
are separated by a smaller distance than the original pair C and E.
The remaining section in between is now fitted in under the
precondition of length conservation. After construction the new
configuration is checked for violation of the topological constraints
and bending energy.  

Naturally, also a backward move has to be
possible that pushes back the region CE into a lesser bend
conformation by creating or enhancing undulations in the region AC.
This process is performed if the random distance $\delta$ is negative
and is achieved in the following way. We proceed in a reverse fashion
by first choosing points B' and C' and reducing their distance by
$\delta$. The same is performed along the tails until the final point
of the region AC is reached. The result is a distance between points C
and E that is now larger than the original distance between the points
C' and E' and thus a section CE that is pulled back from the void to a
straighter configuration and enhanced undulations along the left and
right tails.  Note that exclusive performance of this move will leave
the polymer ends unchanged. In combination with the other moves
however, it allows for a longitudinal motion of the ends by
effectively manipulating the undulations along the encaged contour.

After having constructed a novel trial move for a Monte-Carlo
simulation it is of crucial importance to check if it guarantees a
conversion of the simulation to equilibrium. While it is know that
this can be achieved by the balance condition \cite{manousiouthakis99}
it is usually more convenient to check the stricter condition of
detailed balance \cite{frenkel}. It requires that in equilibrium for
every pair of configurations $m$ and $n$ the moves from $m$ to $n$
given by $P^0_m P^{\mathrm{move}}_{m \to n} P^{\mathrm{acc}}_{m \to
  n}$ equal the number of reverse moves $P^0_n P^{\mathrm{move}}_{n
  \to m} P^{\mathrm{acc}}_{n \to m}$.  Here, $P^0$ is the Boltzmann
weight of a configuration, $P^{\mathrm{move}}$ is the {\it a priori}
probability to select a certain move and $P^{\mathrm{acc}}$ is the
probability that this move is accepted. For the move introduced
above, every move is characterized by the two random numbers
$R_{\mathrm{loc}}$ and $R_{\delta}$ and its backward move is simply
obtained by changing the sign of $\delta$. Consequently, the {\it a
  priori} probabilities of every move and its corresponding backward move cancel. The
acceptance probability is zero for both move and backward move if
topological constraints are violated. If the topology is conserved the
move is accepted according to change in bending energy and thus
according to the ratio of Boltzmann factors and consequently the
condition of detailed balance is guaranteed.

Finally the fourth trial move, a global ``breathing'' move, mimics a
global retraction of the polymer along its contour to enable the
exploration of void spaces. This move causes a pronounced axial motion
of the polymer ends and is schematically depicted in Fig.
\ref{fig:moves} (d). Again, by a random number $R_\mathrm{loc}$ a
section CE is randomly chosen in which the local curvature is to be
manipulated. Either the bending of CE is enhanced resulting in a
global retraction of the remaining tail sections of the polymer to the
left and right hand side of CE, or the bending of CE is diminished
which is achieved by pushing out the remaining sections.  As the
manipulation of the tail sections is always to occur along the
polymer's contour, bending of CE causes axial motion of both ends
towards the polymer's center and straightening out of CE causes end
motion away from the center. In detail the new configuration in the
former case is constructed by choosing a new C' by reducing the
distance between C and E by a small random $\delta$ as above. B' is
then found at that point where a radius $L/N$ intersects with the old
polymer contour. All other points are chosen accordingly proceeding
towards the polymer's ends. The back move is obtained by extending the
section CE by $\delta$ and fixing the direction of the tail segments
by demanding that they pass through the joint points of the old
contour \footnote{E.g. in going back from C' to C, the new point B is
  found by choosing the direction of the new tangent BC to pass
  through B'}. Thereby, again the exact backward move to any given move is
simply obtained by inverting the sign of $\delta$. Consequently, the
same reasoning as above also proves detailed balance.

We validated this particular choice of Monte-Carlo moves in a
simulation of a single free polymer, where we compared our
observations to established results of the bending distribution of
free polymers, tangent-tangent correlation function and end-to-end
distribution function \cite{wilhelm96}. All results presented in the following were
obtained as a combined ensemble and time average. Ensemble averaging was
performed over a large number of initial obstacle fluctuation center
distributions.  Additionally, for every initial obstacle distribution
several initial distributions for the probe filament were chosen. This
can also be seen as averaging over different topologies. After initial
equilibration, observables were monitored and averaged for the
remainder of the simulation time thereby averaging over all
statistically allowed configurations in a fixed topology.

\subsection{Simulation Results}
\begin{figure}[htb]
\begin{center}
\includegraphics[width=0.7\columnwidth]{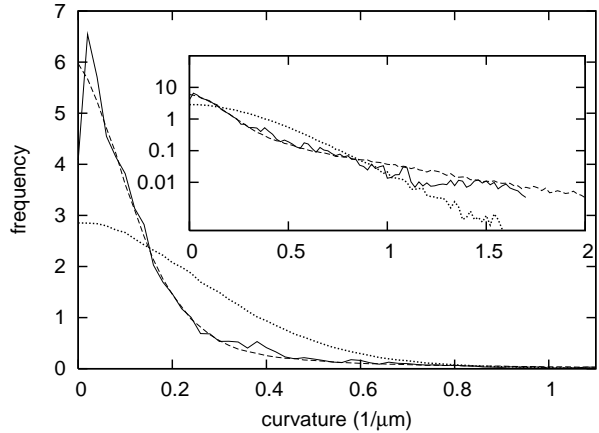}
\caption{Curvature distribution of confinement tube contours as
  obtained from Monte-Carlo simulations (dashed line) and from
  experiments (full line) \cite{romanowska09}.  The
  curvature distribution of a free filament obtained by the same
  simulation algorithm is plotted for comparison (dotted line). The
  inset shows the same data in a semilogarithmic plot. \label{fig:tube}} 
\end{center}
\end{figure}

In a first step we characterized the conformation of the confinement
tubes - a quantity that is also accessible by fluorescence microscopy
and therefore allows for a comparison to experimental data. To this
end we determine the tube contour, i.e. the backbone of the area to
which the probe filament is confined, by averaging over the contour of
the probe polymer in its cage of point obstacles over the evolution.
From the resulting contour we determine a curvature distribution
$P(\mathcal C)$, where the curvature is defined by locally fitting a
parabola with $y=\mathcal Cx^2/2$ to the contour.  The distribution
obtained after averaging over initial conditions is shown in Fig.
\ref{fig:tube}.

We compare our data to measurements that were obtained by fluorescence
microscopy \cite{romanowska09} of in vitro solutions of
rhodamine-phalloidin labeled F-actin on a minute time scale
\footnote{The experimental data was obtained from measurements at
  several concentrations. While the observed effect is in principle
  dependent on concentration as we will discuss below, simulations
  show no significant dependence in the experimental range of actin
  concentration. We thus chose to combine the data of different
  concentrations for the sake of a smaller statistical error.}. The
good agreement of the data is evidence that our simulation approach
provides a reliable representation of the physical system under
consideration.

In comparison to the curvature distribution obtained by the same
algorithm for free filaments, two distinctive differences emerge.
While the free filament distribution has to be Gaussian as explained
in Eq. \ref{eq:distr_free}, the distribution function of the tube
contours features a pronounced exponential tail. As an exponential
decays much slower as a Gaussian towards high values, this signifies
that highly bend filaments are much more frequent. Also for the
occurrence of small curvatures a strong increase in probability
compared to the case of free filaments can be observed. The form of
the distribution, however, remains Gaussian, making the difference
rather quantitative. It is obvious that relative to free filaments,
probability is both shifted to smaller and larger curvatures at the
cost of medium curvatures. This reflects the visual observation that
tube contours are on average straighter than free polymers but also
feature distinctive strongly bend sections. The first feature, i.e.
the increase of small local bendings, obviously results only from the
averaging procedure carried out in determining the tube backbone. The
averaging over all topologically allowed polymer conformations within
the tube integrates out fluctuations of small wavelength (small radii
of curvature) to obtain a larger radius of curvature for the
coarse-grained tube contour. The increase in highly bend filaments on
the exponential tail of the curvature distribution is far less
obvious. To avoid the complications of coarse-graining related to the
tube contour, we turn to a different observable. The curvature
distribution of the encaged filament itself is not as easily
accessible to experiments and thus does not allow for verification,
but it allows a direct comparison to free filaments. In particular,
this is the case for solutions with negligible excluded volume, where
the bending distribution obtained by standard statistical mechanics
should be identical.

\begin{figure}[htb]
\begin{center}
\includegraphics[width=0.7\columnwidth]{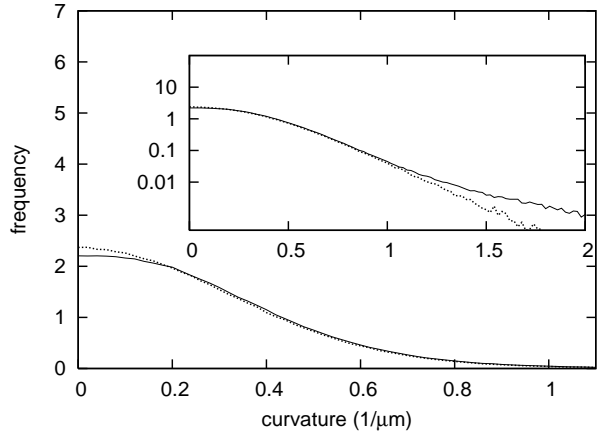}
\caption{Curvature distribution of encaged filaments obtained from
  Monte-Carlo simulations (solid line) compared to the distribution of
  a free filament (dotted line).  The inset shows the same data in a
  semilogarithmic plot. \label{fig:filaments}}
\end{center}
\end{figure}

We recorded snapshots of the probe polymer during the evolution and
analyzed these for their curvature as explained above to obtain the
curvature distribution of confined filaments depicted in
Fig. \ref{fig:filaments}. 
Now the distribution at small and medium curvatures remains largely
unaffected as compared to free filaments. However, the pronounced
exponential tail at high curvatures is still observed. These features
were observed for networks composed of polymers of different
persistence length. At a first glance this behavior seems to
contradict general concepts of statistical mechanics that do not
predict any effect of topological constraints in a system without
excluded volume. We will explain in the following how this conflict is
resolved.

\section{Thermodynamic Interpretation} \label{sec:theory}

As discussed in Sec. \ref{sec:system}, the probe filament is confined
to its tube during the time window that is relevant to many biological
processes and that is the observation frame for most experimental
measurements as well. Clearly, our simulations have also been tailored
to represent this intermediate time scale as point-obstacle centers
are fixed and large scale reptation is beyond simulation time.
Therefore all observations made on this time scale are crucially
influenced by the tube's properties and we will discuss the
implications for the obtained averages and trace back the results for
the deviation of the curvature distribution from the free filament
case. To this end, we will first of all recapitulate some general
notions on standard thermodynamic averaging, then pinpoint differences
to averaging procedures in the tube model and finally present
additional simulations to corroborate our findings. To facilitate our
discussion we restrict ourselves in the following to the case of a
probe filament in a two-dimensional array of fixed point-like
obstacles. This simplified system has the same general characteristic
as a polymer confined to a tube in a network but considerably less
degrees of freedom. 

\subsection{Ensemble Average - Time Average}

If one is faced with the problem to calculate averages for a
statistical system there are in general two different possibilities:
an ensemble average and a time average that will yield the same result
if the system is ergodic. An ensemble average in the system under
consideration could be realized by drawing a large number of allowed
configuration of the complete network and weighing them by the
corresponding Boltzmann factor. As there is no interaction between
polymer and obstacles and in the absence of excluded volume no initial
configuration can be rejected due to hard-core exclusion, the only
contribution to the Boltzmann factor is the bending term of the
worm-like chain (Eq. \ref{eq:hamilton_free}) and the average obtained
has to equal the case of free polymers. Due to the absence of excluded
volume the polymer is not able to see the obstacles and any probe
polymer inserted into the network has zero chance of overlap or
rejection. The concept of a time average, on the contrary, would be to
start from one initial probe polymer configuration and monitor the
following time evolution. As soon as the probe polymer has explored
every point in phase space, the obtained average equals the ensemble
average. In the obstacle system however, the time to fulfill this
requirement is exceedingly long. Points that might be very close to
each other in phase space can be very far apart in terms of transition
time. This is due to the fact that the topological constraints that
the obstacles impose, partition the phase space into a multitude of
areas that are not directly connected. Consequently, the probe
filament can only traverse a point obstacle by completely reptating
back and forth. Due to the immense number of different topologies and
the slow reptation time scale (see Sec.  \ref{sec:system}) a complete
time average is not only out of the scope of simulations and experiments
but also irrelevant to biological processes.

\subsection{Partitioned Averaging}

It can now be tried to substitute the infeasible sampling of phase
space by means of reptation of a single test polymer by a large number
of samples with different initial conditions. Here, initial
configurations of the test polymer are drawn from the free polymer
distribution and randomly placed into the obstacle network. Different
topologies emerge as the same obstacle could be at the left or right
of the test polymer. Starting from these initial conditions the
polymer's configurations are now sampled employing a Markovian
Monte-Carlo dynamics respecting the topological constraints imposed by
the neighboring obstacles. Such a procedure corresponds to a
partitioning of phase space into sections with`different topologies",
i.e. areas that are not directly connected.  One could therefore
argue, that an average containing all possible topologies should also
hold the same results as a complete time average or averaging of a
free polymer.  However, this argument can only be valid if the
partitions of phase space are {\it disjunct}. Otherwise, if these
partitions overlap, the averaging procedure will put a higher or
smaller weight on some microstates.  The curvature distribution
obtained by the experimental and simulational averaging procedure can
thus only be expected to equal the free polymer case, if it is
guaranteed that during the observation time the topology and hence the
partitioning of phase space remains unaltered for every test polymer.
Processes that can modify the topology are for instance reptation or
``breathing" of the polymer. One mutual feature of these processes is
that they cause motion of the polymer ends tangentially along the
contour. Therefore obstacles can switch their topology with respect to
the test polymer, e.g. an obstacle initially left of the test polymer
can end up being on the right side after the polymer's end has moved
back and forth.  The requirement of strict disjunct partitioning of phase
space would thus essentially amount to the constraint of keeping the
polymer's ends fixed which is evidently not the case in the actual
physical system.

We, therefore, conclude that the polymer dynamics inside the
confinement tubes are {\it metrically transitive} due to their
characteristic features as breathing and reptation that change the
topology in the network array. The topological partitioning is {\it thus} not
maintained under the dynamic evolution. Therefore the resulting
averages and distribution functions have not necessarily to equal the
corresponding results for free polymers. This holds on intermediate
time scales before large scale reptation sets in, which are the time
scales of experimental observation. In the long time
limit, however, when the single polymers of the ensemble have been able
to explore larger parts of the phase space beyond their confinement
tubes, free filament distributions should be recovered. Consequently,
the observed non-equilibrium distribution functions do not violate
thermodynamic requirements as they are transient. However, the time
needed for total equilibration is so long, that it is not reached on
any applicable time scale.

After we have shown, how transient non-equilibrium distribution
functions can arise on intermediate time scales even in the absence of
excluded volume, we will use additional simulations to clarify the
physical origins of highly bend filaments.

\begin{figure}[htb]
\begin{center}
\includegraphics[width=0.7\columnwidth]{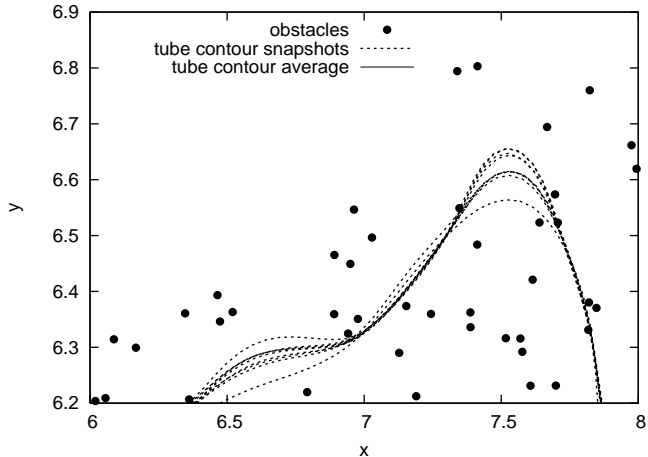}
\caption{A typical network configuration where transient entropic
  trapping occurs when the probe filament explores a void space by
  high bending thereby realizing an entropic gain. \label{fig:snapshot}}
\end{center}
\end{figure}

\subsection{Additional Simulations - Entropic Trapping}

\begin{figure}[htb]
\begin{center}
\includegraphics[width=0.7\columnwidth]{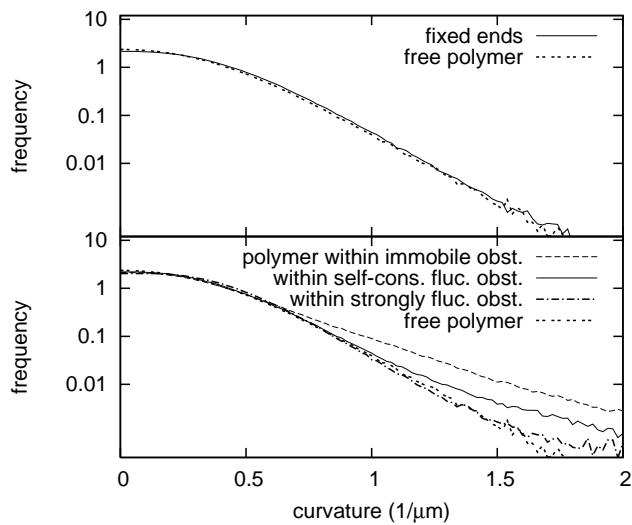}
\caption{Curvature distributions compared to free filaments (dotted
  line): {\it (top)} metrically intransitive system with fixed ends
  (solid line), {\it (bottom)} probe filament in obstacles of
  different fluctuation strength. \label{fig:moredistr}}
\end{center}
\end{figure}

To this end we have conducted further simulations of the simplified
system of fixed obstacles. Note, that this work does not apply to a
network of F-Actin as represented by the simulations with
self-consistently fluctuating obstacles. It merely serves as an model
system for our considerations on the averaging procedures. First of
all, we have checked if a system where dynamics have chosen to be
metrically intransitive faithfully reproduces the distribution
functions of free filaments. This was achieved by running simulations
where the filament's Monte-Carlo moves are restricted to ``flip'' and
``trade-off'' moves. This choice ensures that the polymer's ends do
not move axially. As obstacles remain fixed, it is guaranteed that the
topological partitioning {\it cannot} change. The resulting distribution
function indeed reproduces the case of free filaments (see Fig.
\ref{fig:moredistr} (top)). Furthermore, we have identified the
physical origins of the highly bend parts of the test polymer. It
turns out that high curvatures occur always at local initial
topologies where a test polymer can protrude into a large void space
in the obstacle array (see Fig. \ref{fig:snapshot}).  The initial
conformation of the test polymer already has a curvature that
facilitates a further bending into a large void part in the network.
Hereby the system realizes a higher entropy by bending harder than the
equilibrium curvature distribution.  These events are rare but they
dominate the tail of the curvature distribution.  Apparently, the
polymer is trapped in these entropically favorable configurations on
the time scale of observation.  This behavior bears some similarity to
``entropic trapping'' observed for flexible polymers in random
environments \cite{baumgartner87,cates88,edwards88,sommer97}. Note in particular, that
these conformations also result in a pulling back of the polymer ends
and thus a change of phase space partition.  Hence, this observation
does not only clarify the physical cause of high bendings but also
proves according to the argumentation above that a curvature
distribution different from the free polymer distribution does not
violate statistical mechanics. 

Furthermore, we have investigated how
these special conformations are realized as a function of the
fluctuation amplitude of the obstacles. For both self-consistently
fluctuating and immobile obstacles an exponential tail is visible in
the curvature distribution. This is also the case for obstacles
fluctuating with a higher amplitude as in the self-consistent case.
The weight on the high curvature tail is highest for immobile
obstacles and decreases with increasing fluctuation amplitude. This is
consistent with the explanation for the high bendings provided above.
As the effective size of void spaces is diminished with larger
obstacle fluctuation the effect decreases. In the limiting case of
very large fluctuations network obstacles become delocalized, the
system is reduced to a gas and the curvature distribution of a free
polymer will be recovered.

\section{Conclusion} \label{sec:conclusion} 

We have investigated the curvature distribution functions in entangled
networks of semiflexible polymers. To this end we developed an
approach to simulate single probe filaments in an entangled network of
semi-flexible polymers by a self-consistent reduction to a
two-dimensional setup corresponding e.g. to the focal plane of a
microscope to allow for comparison for experimental data. This
Monte-Carlo simulations were particularly designed to mimic the real
polymer dynamics on intermediate time scales by allowing for an
effective exploration of network void spaces by breathing and short
scale reptation. The simulations provide data on curvature
distributions for tube contours that agree well with fluorescence
microscope measurements on F-actin solutions \cite{romanowska09}.
Furthermore, they permit to observe curvature distributions of single
confined filaments. These distributions feature an unexpectedly high
weight on highly bend filaments that is traced back to transient
entropic trapping in network void spaces. The fact that the
equilibrium distribution of free polymers is not recovered even in the
absence of excluded volume, is shown to be an immanent feature of the
polymer dynamics in a disordered environment on intermediate time
scales below large scale reptation. The fact that this regime is best
described by the tube model - a non-equilibrium concept - explains
that a treatment in terms of equilibrium thermodynamics is
inappropriate. Consequently, the observation of transient
non-equilibrium distribution functions is a generic effect observed
for all measurements on time scales both relevant to experiments and
feasible for simulations. These findings provide insight into the
conformation of confined polymers and can e.g. prove useful for
further analysis of reptation or emerging collective macroscopic
properties.

\bigskip

{\bf Acknowledgement:}
  We thank R. Merkel and his group for fruitful cooperation and J.U.
  Sommer for discussions on ``entropic trapping''.  We acknowledge
  support from the DFG through SFB 486, from the German Excellence
  Initiative via the NIM program and from the Elite Network of Bavaria
  through the NBT program.

\end{document}